\documentclass[letterpaper,aps,preprintnumbers,twocolumn,amsmath,amssymb]{revtex4}

\usepackage{amsmath,amsfonts,amssymb,graphics,graphicx,epsfig,color,times,bbm}
\usepackage{amsthm}

\usepackage{enumerate}

 \newtheorem{theorem}{Theorem}
 \newtheorem{proposition}{Proposition}

\newcommand*{\1}{\mathbbm{1}}

\newcommand*{\cH}{\mathcal{H}}

\newcommand*{\cV}{\mathcal{V}}

\newcommand*{\tr}{\mathsf{tr}}

\newcommand*{\E}{\mathsf{E}}
\newcommand*{\tE}{\tilde{\mathsf{E}}}

\newcommand*{\F}{\mathsf{F}}

\begin{document}

\title{On Exchangeable Continuous Variable Systems}

\author{Robert K\"onig$^{1}$ and Michael M. Wolf$^{2,3}$}

\affiliation{\vspace{1ex} $^1$Institute for Quantum Information, Caltech, Pasadena, USA\\
$^2$Max-Planck-Institut f\"ur Quantenoptik, Garching, Germany\\
$^3$Niels Bohr Institute, Copenhagen, Denmark}


\date{\today}


\begin{abstract}
We investigate permutation-invariant continuous variable quantum states and their covariance matrices. We provide a  complete characterization of the latter with respect to permutation-invariance, exchangeability and representing convex combinations of tensor power states. On the level of the respective density operators this leads to necessary criteria for all these properties which become necessary and sufficient for Gaussian states. For these we use the derived results to provide de Finetti-type theorems for various distance measures.
\end{abstract}

\maketitle

\section{Introduction}
A $k$-partite density operator on~$(\mathbb{C}^d)^{\otimes k}$ is called {\em $n$-exchangeable} if it is the partial trace of a permutation-invariant state on $(\mathbb{C}^d)^{\otimes n}$.
According to the quantum de Finetti theorem, such states can be approximated  by convex combinations of $k$-fold product states with error bounded by $O(\frac{dk}{n})$. This result has various applications in quantum information theory (e.g., in security proofs for cryptographic protocols~\cite{Ren05,Hor07} or the justification of mean-field approaches~\cite{raggiowerner}) and relatives (the ``monogamy'' of entanglement (cf.~\cite{VerstraeteOsborne,WVCfrustration}) or the appearance of a local classical description under symmetries~\cite{Wer89b,WVC2,TerDohSch}).

Unfortunately, such de Finetti-type statements for finitely exchangeable states are no longer true when the Hilbert space $\mathbb{C}^d$ of an individual (``local'') system is replaced by an infinite-dimensional Hilbert space---as it happens for instance when dealing with fields of light. Indeed, as shown in~\cite{chrkoemire06}, there are $n$-exchangeable states on $(\mathbb{C}^d)^{\otimes k}$ whose distance from any convex combination of product states is at least $\Omega(\frac{d}{n})$. This outrules the possibility of there being a de Finetti theorem for all finitely exchangeable states on infinite-dimensional Hilbert spaces.

It is, however, still possible to find de Finetti-type results for infinite-dimensional systems by considering  restricted classes of states or observables~\footnote{See~\cite{christandltoner07,barrettleifer07} for a different direction of generalization of the basic de Finetti claim for quantum states.}. An example are the {\em coherent cat-states} for which a de Finetti theorem was obtained by D'Cruz, Osborne and Schack~\cite{CruzOsborneSchack06}; an ``exponential'' version thereof is provided in~\cite{koemit07}.
In the present work we will first restrict the set of observables and consider covariance matrices of canonical field operators (such as position and momentum or the quadratures of a field of light). In the second part we will then use these results in order to derive de Finetti-type theorems for a restricted class of states, namely {\em Gaussian states}. These play an important role in quantum optics, as coherent, squeezed and their thermal states are all Gaussian and remain so under the action of quadratic Hamiltonians and homodyne measurements.

Symmetry and exchangeability for infinite-dimensional systems are defined analogously to the finite-dimensional case: We substitute the local Hilbert spaces~$\mathbb{C}^d$ by the Hilbert space $\cH_d:=\cH^{\otimes d}$ of $d$~harmonic oscillators or {\em modes} with $\cH\cong{\cal L}^2$ being infinite-dimensional. The symmetric group $\mathsf{S}_n$ acts on the $n$-fold tensor product $\cH_d^{\otimes n}$ by permuting the factors. A state $\rho$ on $\cH_d^{\otimes n}$ is called {\em symmetric} if it is invariant under this action, i.e., if  $\pi\rho\pi^\dagger=\rho\ \textrm{ for all }\pi\in\mathsf{S}_n$. States $\tr_{\cH_d^{\otimes n-k}}\rho$  on~$\cH_d^{\otimes k}$ obtained by tracing out $n-k$~``local'' systems of a symmetric state $\rho$ on $\cH_d^{\otimes n}$ are called {\em $n$-exchangeable}. {\em Infinitely exchangeable states} on $\cH_d^{\otimes k}$ are those that are $n$-exchangeable for any $n\geq k$; according to the infinite-exchangeability de Finetti theorem~\cite{HudMoo76,stormer69,cavesfuchsschack}, which also holds on infinite-dimensional Hilbert spaces, these are all states of the form $\int \tau^{\otimes k}dm(\tau)$, i.e., convex combinations of $k$-fold product states $\tau^{\otimes k}$, where $\tau$ are states on the single system $\cH_d$. We call such convex combinations {\em power-states}.

We will provide some prerequisites in the next section and then give a detailed discussion of permutation-invariant covariance matrices. This will yield necessary criteria for the respective density matrices which become sufficient as well when considering Gaussian states. De Finetti theorems for the latter are then provided in the second part.

\section{Preliminaries}
This section will introduce the notation and recall basic results (cf. \cite{hol82}) on covariance matrices and the corresponding continuous variable states.
Consider a system of $n$ sites, each of which with $d$ canonical degrees of freedom (modes). Such a system can be described using a symplectic
vector space, the \emph{phase space}, $\cV\cong\mathbb{R}^{2d}\otimes\mathbb{R}^n$ of dimension
$2d\cdot n$. We choose canonical coordinates such that the {\em symplectic
  matrix} $\sigma$
has the  form
\begin{align}
\sigma=\left(\begin{matrix} 0_d
  & -\1_d\\
\1_d& 0_d\end{matrix}\right)\otimes\1_n\ ,  \label{eq:symplecticmatrix}
\end{align}
where $0_d$ and $\1_d$ are the $d\times d$-zero and identity matrix respectively. Occasionally, we will denote with a slight abuse of notation the symplectic matrix for a subsystem (less than $n$ sites) by $\sigma$ as well. The block structure in Eq.(\ref{eq:symplecticmatrix})
reflects the grouping into position- and momentum
coordinates. The corresponding quantum system with Hilbert space $\cH_d^{\otimes n}\cong({\cal L}^2)^{\otimes dn}$
 is characterized by
 canonical position- and momentum operators
 $\{R^\alpha_i\}_{\alpha=1}^{2d}$ for each site
 $i=1,\ldots,n$; here $\{R^\alpha_i,R^{d+\alpha}_i\}$
for $1\leq \alpha\leq d$ are the position- and momentum-operator of
the $\alpha$-th mode of the $i$-th site, acting
non-trivially only on the $i$-th factor of the $n$-fold tensor product
space $\cH_d^{\otimes n}$. These operators satisfy the
canonical commutation relations (CCR)
\begin{align}\label{eq:canonicalcommutation}
i[R_k^\alpha,R_l^\beta] = \sigma_{(\alpha, k),(\beta, l)}\1\ .
\end{align}
Real linear transformations $S:\cV\rightarrow\cV$ which preserve the CCR (i.e., $S\sigma S^T=\sigma$) are called \emph{symplectic}. Affine transformations are obtained from the  unitary {\em Weyl operators}
$W(\xi):=e^{i\xi^T\sigma R}$ \footnote{Here we assume, as usual, the Weyl system to be irreducible and strongly continuous.} which give rise to displacements:
\begin{align}\label{eq:displacement}
W(\xi)R^\alpha_i W(\xi)^\dagger=R^\alpha_i+\xi^\alpha_i\1\ .
\end{align}
These only alter the first moments of a state, collected in the {\em displacement vector} $D:=\tr(\rho R)$, while keeping untouched the covariance matrix (CM) $\Gamma$ with entries
$$ \Gamma_{(\alpha, i),(\beta,
  j)}=\tr\left(\{R^\alpha_i-D^\alpha_i\cdot\1,R^\beta_j-D^\beta_j\cdot\1\}_+\rho\right)\ .$$

Since the Weyl operators act locally in the sense that they factorize with respect to local
systems we can w.l.o.g. typically assume $D=0$. Positivity of the density matrices implies that any CM has to satisfy $\Gamma\geq i\sigma$. Conversely, every real symmetric matrix fulfilling this requirement is a valid CM. In particular, there always exists a \emph{Gaussian state} which is, up to displacements, completely characterized by this CM.

\section{Characterization of covariance matrices}

We are now ready to characterize covariance matrices with respect to permutation-invariance, exchangeability and power-state property. Clearly, we could express all results below in terms of Gaussian states. However, we want to logically separate the treatment of CMs, since it is neither necessary for the state to be Gaussian nor does the corresponding density matrix have to share the full symmetry of the CM.

We will say that a CM is a power-state CM if it corresponds to a mixture of states with product CMs of the form~$\gamma\otimes\1_n$. Similarly, we say that it is $n$-exchangeable if it is a sub-block of a permutation-invariant CM of $n$ sites, and we call it separable if it corresponds to a mixture of states with product CMs of the form $\bigoplus_{i=1}^n\gamma_i$.

The main ingredient of our analysis is the following one-to-one correspondence between permutation-invariant CMs and bipartite product CMs:
\begin{proposition}[Symmetric CMs]\label{prop:symCM}
Let $X$ be any real orthogonal $n\times n$ matrix for which $X_{i,1}=1/\sqrt{n}$ for all $i=1,\ldots,n$. Then $S=\1_{2d}\otimes X$ is a symplectic transformation on $\cV=\mathbb{R}^{2d}\otimes\mathbb{R}^n$ which gives rise to a one-to-one mapping between permutation-invariant CMs $\Gamma$ on $\cV$ and pairs of CMs $\E,\F$ on $\mathbb{R}^{2d}$ such that
\begin{align}\label{eq:gammaefdecomp}
\Gamma=\Gamma(\E,\F)=S(\E\oplus \bigoplus_{i=2}^{n} \F)S^T\ .
\end{align}
\end{proposition}
\begin{proof}
Evidently, $S$ is a symplectic transformation and therefore maps CMs onto CMs. In fact, it is a passive transformation, i.e., number preserving.
We first show that every pair of CMs $\E,\F$ leads to a permutation-invariant~$\Gamma$. For this we have to show that the entries of the latter are site-independent in the sense that
\begin{align}\label{eq:abdef}
\Gamma_{(\alpha, i),(\beta, i)}&=A_{\alpha,\beta}\ \textrm{,\ \ \ }
\Gamma_{(\alpha, i),(\beta, j)}&=B_{\alpha,\beta}\ \textrm{ for } i\neq j\ .
\end{align}
The symmetric $2d\times 2d$ matrices $A$ and $B$ encode the local (on-site) and non-local (inter-site) correlations,
respectively. With some abuse of Dirac notation  Eq.(\ref{eq:gammaefdecomp}) can be written as
\begin{equation}
\Gamma=E\otimes X|1\rangle\langle 1|X^T+F\otimes\big(\1-X|1\rangle\langle 1|X^T\big).
\end{equation}
Exploiting the condition on the first column of $X$ this leads to
\begin{equation}
\Gamma_{(\alpha, i),(\beta, j)}=\frac1n \E_{\alpha,\beta}+(\delta_{i,j}-\frac1n )\F_{\alpha,\beta},
\end{equation}
which is indeed permutation-invariant. This immediately implies that every permutation-invariant CM characterized by $A,B$ can be obtained in the above way by choosing \begin{align}\label{eq:efabdef}
\E=A+(n-1)B\qquad \textrm{ and }\qquad \F=A-B\ .
\end{align}
\end{proof}
We will now use this representation in order to characterize power-state and exchangeable CMs:
\begin{proposition}[Power-state CMs]\label{prop:power}
Let $\Gamma(\E,\F)$ be a permutation-invariant CM on $\mathbb{R}^{2d}\otimes\mathbb{R}^n$. Then the following are equivalent: (i) $\Gamma$ is a power-state CM, (ii) $\Gamma$ is a separable CM, and (iii) there exists a CM $\gamma$ of $d$ modes such that $\E\geq\gamma$ and $\F\geq\gamma$.
\end{proposition}
\begin{proof}
Clearly, (i) implies (ii). The remaining proof closely follows the derivation of
the separability criterion for Gaussian states in~\cite{WW01}. There it is shown that for every separable CM $\Gamma$ there exist CMs $\gamma_i$ such that $\Gamma\geq\bigoplus_{i=1}^n\gamma_i$. Averaging over all permutations we obtain $\Gamma\geq\gamma\otimes\1_n$ with $\gamma=\frac1n \sum_i\gamma_i$. Applying the symplectic transformation of Prop.\ref{prop:symCM} to this inequality leads to $\E,\F\geq\gamma$ (since the r.h.s. has $\E=\F=\gamma$) an thus (iii) is equivalent to (ii). Finally, again following~\cite{WW01}, $\Gamma\geq\gamma\otimes\1_n$ implies that $\Gamma$ is the CM of a mixture of states with CM $\gamma\otimes\1_n$.
\end{proof}
\begin{proposition}[Exchangeable CMs]\label{lem:symmgaussian}
Let $\Gamma$ be a permutation-invariant CM on $\mathbb{R}^{2d}\otimes\mathbb{R}^k$, $k\leq n$. With the notation from Prop.\ref{prop:symCM} $\Gamma$ is  $n$-exchangeable  if and only if $A+(n-1)B\geq i\sigma$ or equivalently $\E=\frac{k}n \tilde\E+(1-\frac{k}n)\F$ with $\tilde\E$ being any CM of $d$ modes.
\end{proposition}
This follows in a straightforward way from Prop.\ref{prop:symCM} and the relation Eq.(\ref{eq:efabdef}).
It is remarkable that Prop.~\ref{lem:symmgaussian} gives a simple
criterion for deciding $n$-exchangeability for any CM. Note that such a criterion is not known for general
exchangeable states on finite-dimensional Hilbert spaces
(cf.~\cite{fannesvandenplas06}).

We also point out that according to Prop.~\ref{lem:symmgaussian},
$\Gamma$ admits a permutation-invariant super-system of~$n$ sites for all $n\geq k$ if and only if $\E\geq \F$; this is a
stronger condition than the one for
 power-state CMs.

\section{De Finetti theorems for Gaussian states}
We now step back from phase space to Hilbert space and derive upper bounds on the distance between exchangeable Gaussian states and the set of power-states. Note that the conditions for a Gaussian state to be exchangeable or a power-state are precisely those given in the previous section. We begin with the single-mode case ($d=1$) and recall that the trace norm ($L^1$-norm) is defined by $\|A\|_1=\tr\sqrt{A^\dagger A}$.
\begin{theorem}[Single-mode de Finetti]\label{thm:gaussiandefinetti}
Let $\rho_{\Gamma(\E,\F)}$ be an $n$-exchangeable Gaussian  density matrix describing a permutation-invariant $k\leq n$ mode system. Define $\E'=\frac{n}{n-k}\E$.
Then $\rho_{\Gamma(\E',\F)}$ is a Gaussian power-state, and
\begin{align*}
\big\|\rho_{\Gamma(\E,\F)}-\rho_{\Gamma(\E',\F)}\big\|_1\leq
\Big(\frac{n}k-\frac12\Big)^{-1}.
\end{align*}
\end{theorem}
\begin{proof}
The matrix $\E'$ defines a valid CM, since multiplying by a factor
$\lambda:=\frac{n}{n-k}\geq 1$ preserves the condition $\E\geq i\sigma$.  Using $n$-exchangeability and the
CM~$\tE$ appearing in Prop.\ref{lem:symmgaussian} we can write
$\E'=\frac{k}{n-k}\tE+\F$.  Observe that $\tE$ is a positive semi-definite
operator, which is implied by $\tE\geq i\sigma$ (via complex conjugation and averaging). We conclude that
$\E'\geq \F$. According to Prop.\ref{prop:power},  this implies that $\rho_{\Gamma(\E',\F)}$ is indeed a Gaussian power-state.

To compute
the distance between $\rho_{\Gamma(\E',\F)}$ and
$\rho_{\Gamma(\E,F)}$, we
use the tensor product form of the states
in the basis of Prop.\ref{prop:symCM} together with the stability property
$\|\rho\otimes\sigma-\rho'\otimes\sigma\|=\|\rho-\rho'\|$ of the
trace norm, so that
\begin{equation}
\big\|\rho_{\Gamma(\E,\F)}-\rho_{\Gamma(\E',\F)}\big\|_1=\big\|\rho_{\E}-\rho_{\E'}\big\|_1\ .
\end{equation}
To evaluate this further we use that there is a canonical basis where $\E=s\1_2$ for some $s\geq 1$ and thus $\E'=\lambda s\1_2$. The corresponding density matrices are now simultaneously diagonal in Fock state basis (labeled by $\ell=0,\ldots,\infty$) with eigenvalues $\mu_\ell(s):=2 (s-1)^\ell/(s+1)^{\ell+1}$ and $\mu_\ell(\lambda s)$ respectively. Hence, we proceed with \footnote{Eq.(\ref{eq:normsum}) is evaluated by first noting that $\mu_\ell(s)$ and $\mu_\ell(\lambda s)$ are normalized geometric series such that one is larger than the other up to some specific~$\ell$. In this way the absolute value can be expressed as a difference of two geometric series.}
\begin{eqnarray}\label{eq:normsum}
\big\|\rho_{\E}-\rho_{\E'}\big\|_1 &=& \sum_{\ell=0}^\infty \big|\mu_\ell(s)-\mu_\ell(\lambda s)\big|\\
&=& \max_{\ell\in\mathbb{N}} 2\left[\Big(\frac{\lambda s-1}{\lambda s+1}\Big)^{\ell+1}-\Big(\frac{ s-1}{ s+1}\Big)^{\ell+1}\right]\nonumber.
\end{eqnarray}
In order to obtain a bound which is independent of $s$ we consider the worst case, i.e., we take the supremum over $s\geq 1$ which is achieved for $s=1$ \footnote{For this we need to evaluate the expression at the optimal $\ell$ which is one of the two integers next to $$\left(\log\Big(\frac{\log\nu(s)}{\log\nu(\lambda s)}\Big)/\log\big(\frac{\nu(\lambda s)}{\nu(s)}\big)\right)-1,$$ with $\nu(s):=(s-1)/(s+1)$.} for which $\ell=0$ attains the maximum so that finally
$$\big\|\rho_{\E}-\rho_{\E'}\big\|_1\leq 2 \frac{\lambda-1}{\lambda+1}=\Big(\frac{n}k-\frac12\Big)^{-1}.$$
\end{proof}

Let us now consider Gaussian systems with several modes per site ($d\geq 1$). In principle we could follow the same route as above. However, computing the trace-norm distance directly appears to become cumbersome due to its non-additivity with respect to tensor products. For this reason we choose two different figures of merit for measuring the distance to the set of power-states: the relative entropy\cite{OhyaPetz} $S(\rho,\rho')=\tr\big[\rho(\log\rho-\log\rho')\big]$ and the fidelity $F(\rho,\rho')=\tr\sqrt{\sqrt{\rho}\rho'\sqrt{\rho}}$. Both have benign behavior under taking tensor products (they are additive and multiplicative, respectively) and both yield bounds for the trace norm distance since $$\frac12||\rho-\rho'||_1^2\leq\left\{\begin{array}{c}
                                        S(\rho,\rho'), \\
                                        2-2F(\rho,\rho')^2
                                      \end{array}\right. .$$

\begin{theorem}[Multi-mode de Finetti]\label{thm:gaussiandefinetti2}
Let $\rho_{\Gamma(\E,\F)}$ be an $n$-exchangeable Gaussian  density matrix describing a permutation-invariant system of $k\leq n$ sites of $d$ modes each.
Then the Gaussian power-state $\rho_{\Gamma(\E',\F)}$ with $\E'=\frac{n}{n-k}\E$  satisfies
\begin{eqnarray}
S\Big(\rho_{\Gamma(\E,\F)},\rho_{\Gamma(\E',\F)}\Big)&\leq&
\log\left(\frac{n-k/2}{n-k}\right)^d,\\
F\Big(\rho_{\Gamma(\E,\F)},\rho_{\Gamma(\E',\F)}\Big)&\geq& \left(\frac{n-k}{n-k/2}\right)^{d/2}\;.
\end{eqnarray}
\end{theorem}
\begin{proof} It is clear from the arguments in Thm.\ref{thm:gaussiandefinetti} that $\Gamma(\E',\F)$ indeed corresponds to a Gaussian power-state. We again exploit that the distance between $\rho_{\Gamma(\E,\F)}$ and $\rho_{\Gamma(\E',\F)}$ equals that between $\rho_{\E}$ an $\rho_{\E'}$. Moreover, we can simultaneously diagonalize the latter two states such that $\rho_{\E}=\bigotimes_{\alpha=1}^d \rho(s_\alpha)$ where each $\rho(s_\alpha)=\sum_{\ell=0}^\infty\mu_\ell(s_\alpha)|\ell\rangle\langle \ell|$ is a thermal Gaussian state with CM $s_\alpha\1_2$ and the $s_\alpha\geq 1$ are the symplectic eigenvalues of $\E$. For $\E'$ we only have to replace $s_\alpha$ by $\lambda s_\alpha$ with $\lambda=n/(n-k)$.  For the relative entropy, summing up the series leads to
\begin{eqnarray}
&&\hspace{-15pt} S\Big(\rho_{\Gamma(\E,\F)},\rho_{\Gamma(\E',\F)}\Big)=\sum_{\alpha=1}^d S\big(\rho(s_\alpha),\rho(\lambda s_\alpha)\big)\\
&=&\frac12\sum_{\alpha=1}^d\Big((s_\alpha-1)\log(s_\alpha-1)-(s_\alpha+1)\log(s_\alpha+1)\nonumber\\
&&+ (s_\alpha+1)\log(\lambda s_\alpha+1)-(s_\alpha-1)\log(\lambda s_\alpha-1)\Big). \nonumber
\end{eqnarray}
 For every $\lambda\geq1$ each summand is a decreasing function in $s_\alpha$ such that the supremum is again obtained for $s_\alpha=1$, i.e.,
$$\sup_{s\geq 1} S\big(\rho(s),\rho(\lambda s)\big)=\log\frac{1+\lambda}2$$ which concludes the proof for the relative entropy upon inserting $\lambda$.
In a similar vein we can evaluate the fidelity:
\begin{eqnarray}
F\Big(\rho_{\Gamma(\E,\F)},\rho_{\Gamma(\E',\F)}\Big)&=&\prod_{\alpha=1}^d F\big(\rho(s_\alpha),\rho(\lambda s_\alpha)\big),\\
F\big(\rho(s),\rho(\lambda s)\big) &=& \sum_{\ell=0}^\infty \sqrt{\mu_\ell(s)\mu_\ell(\lambda s)}\nonumber\\
&&\hspace{-80pt}=\frac2{\sqrt{(s+1)(\lambda s+1)}-\sqrt{(s-1)(\lambda s-1)}}.
\end{eqnarray}
Again the worst case, now the infimum over $s\geq 1$, is attained at $s=1$ which leads to the desired result.
\end{proof}

\section{Conclusions \& Open Problems}
The bounds given in Thm.\ref{thm:gaussiandefinetti} and Thm.\ref{thm:gaussiandefinetti2} are tight for the given class of ansatz states and the proofs allow for any explicitly given CM $\E$ to compute the exact distance to this class. Whether the chosen class is optimal or yields at least an optimal scaling remains open. Similarly, the question for which figure of merit a power-state closest to a given $n$-exchangeable Gaussian state can again be chosen to be Gaussian
remains open as standard arguments, e.g. based on the central limit theorem (CLT) \cite{wolfetal06}, do not immediately apply. Some kind of marriage between the CLT and the de Finetti theorem would also be desirable in order to extend the latter in a reasonable way beyond the class of Gaussian states.

Another interesting direction of future research could be to use more general approximating states than Gaussian power-states in the de Finetti theorem. In this way it might be possible to obtain an exponentially small error as in the
``almost-product'' de Finetti theorem for  finite-dimensional
systems~\cite{Ren05} (see also~\cite{renner-2007,koemit07}). This would be
important for applications to continuous-variable quantum
key distribution, for example~\cite{garciacerf06}.

\acknowledgements

RK would like to thank the Max-Planck-Institute for its hospitality, and Ignacio Cirac and Geza Giedke for helpful discussions. He acknowledges partial support by 
NSA under ARO contract no.~W911NF-05-1-0294 and by NSF under contract no.~PHY-0456720.

\end{document}